\documentclass[12pt]{iopart}

\usepackage{graphicx} 
\begin{document}

\title[Bondi-type accretion in the Reissner--Nordstr\"{o}m--(anti--)de Sitter spacetime]{Bondi-type accretion in the Reissner--Nordstr\"{o}m--(anti--)de Sitter spacetime}

\author{F Ficek}

\address{Instytut Fizyki im. Mariana Smoluchowskiego, Uniwersytet Jagiello\'nski,   \L{}ojasiewicza 11,  30-348 Krak\'ow,  Poland}
\ead{filip.ficek@student.uj.edu.pl}
\vspace{10pt}
\begin{indented}
\item[]September 2015
\end{indented}

\begin{abstract}
In this paper I study stationary, spherically symmetric accretion of fluids onto a charged black hole in the presence of the cosmological constant. For some isothermal equations of state it is possible to obtain analytic solutions. For the case of a radiation fluid I derive a connection between locations of horizons and sonic (critical) points. In specific cases the solutions form closed, binoculars-like trajectories in a phase diagram of the velocity vs. radius.
\end{abstract}


\submitto{\CQG}
\maketitle

\section{Introduction}

In a series of recent papers \cite{karkowski_malec, Mach, mach_malec, mach_sam} Karkowski, Mach and Malec investigated spherically symmetric Bondi-type accretion of perfect fluids in Schwarzschild--(anti--)de Sitter spacetimes. In \cite{karkowski_malec} and \cite{Mach} a more general case of self gravitating fluid is also studied. The primary motivation of these works was cosmological; the authors were mainly trying to determine the way in which the presence of the cosmological constant influences the possible accretion rate.

In this paper I go a step further and consider relativistic Bondi-type accretion \cite{Bondi, Michel} in the Reissner--Nordstr\"{o}m--(anti--)de Sitter spacetime. The motivation behind this choice comes from the fact that the structure of Reissner--Nordstr\"{o}m metrics resembles certain features characteristic for the Kerr solution that could influence accretion \cite{Poisson}. They share similar horizon structure \cite{Israel}. Also, the Penrose diagram for a generic case of Reissner--Nordstr\"{o}m--de Sitter spacetime and the Penrose diagram for Kerr--de Sitter universe are similar \cite{Akcay}. These similarities in causal and horizon structures let one suppose that accretion solutions in both spacetimes may be similar. In fact it is common to treat the Reissner--Nordstr\"{o}m solutions as a toy model for astrophysical black holes \cite{Israel, Poisson, Burko, Dafermos}. Finding general, radially dominated, stationary accretion flows on the Kerr metric seems to be intractable at present (cf.\ \cite{Mendoza1, Mendoza2, Tejeda}), with the exception of the ultra-relativistic potential flows \cite{Petrich, Babichev}. Investigating spherically symmetric solutions in the Reissner--Nordstr\"{o}m--(anti--)de Sitter spacetime seems to be a way to avoid (at least temporarily) the main mathematical difficulties of dealing with metrics that are not spherically symmetric. I believe, however, that steady Bondi-type accretion in the Kerr spacetime and in Reissner--Nordstr\"{o}m-type spacetimes could share some features of common behaviour, at least qualitatively.

Reissner--Nordstr\"{o}m--de Sitter black holes can also be considered in the context of the  anti--de Sitter/conformal field theory correspondence \cite{CFT, instability}. Thermodynamical properties of the Reissner--Nordstr\"{o}m--(anti--)de Sitter black holes are a subject of numerous investigations \cite{termo1, termo2, termo3, dwymiary}. 

Very recently Chaverra and Sarbach presented another analysis of Bondi-type accretion flows on essentially arbitrary, spherically symmetric metrics \cite{Chaverra}. The solutions discussed in this paper fall in their class. Of course, by specializing to a narrower set of Reissner--Nordstr\"{o}m--(anti--)de Sitter spacetimes I was able to obtain much more detailed information about corresponding accretion solutions.

The order of this paper is as follows. In Section 2 I introduce the Reissner--Nordstr\"{o}m--(anti--)de Sitter metric and specify the coordinate system. Section 3 introduces general equations of the flow. I also define the notion of the sonic point there. Section 4 uses results of the previous section in case of different types of isothermal test fluids. I show, that for some of them it is possible to obtain analytic solutions. I plot some of them in a phase diagram of the velocity vs. radius. It shows, that for subrelativistic fluid and negative cosmological constant it is possible to obtain closed, binoculars-like trajectories. Section 5 yields results for polytropic test fluids. Also for them there exist closed trajectories. Section 6 summerizes the paper and compares the obtained results. In the Appendix I prove the relation between locations of the horizons and the sonic points in case of the radiative fluid in Reissner--Nordstr\"{o}m--(anti--)de Sitter spacetime.

\section{Reissner--Nordstr\"{o}m--(anti--)de Sitter spacetime}
In polar coordinates $(t,r,\theta,\phi)$ the metric of the Reissner--Nordstr\"{o}m--(anti--)de Sitter spacetime can be written as \cite{termo3}:
\begin{eqnarray}
\fl ds^2=-\left(1-\frac{2m}{r}+\frac{Q^2}{r^2}-\frac{\Lambda}{3}r^2\right) dt^2+\frac{1}{\left(1-\frac{2m}{r}+\frac{Q^2}{r^2}-\frac{\Lambda}{3}r^2\right)} dr^2+\nonumber\\
+r^2 (d\theta^2+\sin^2 (\theta) d\phi^2),
\label{eqn:metric}
\end{eqnarray}
where $m$ and $Q$ correspond to the mass and the charge of a black hole, and $\Lambda$ is a cosmological constant. This coordinate system is singular at the solutions of the equation
\begin{equation} 
1-\frac{2m}{r}+\frac{Q^2}{r^2}-\frac{\Lambda}{3}r^2=0.
\label{eqn:singular}
\end{equation}
This equation has at most three real, positive roots, corresponding to the Cauchy horizon, the event horizon and the cosmological horizon. The last one exists only for de Sitter spacetimes ($\Lambda>0$). In order to get generic case where these three horizons exist (or two in case of $\Lambda<0$) one imposes constraints on the values of $m$, $Q$ and $\Lambda$. For $\Lambda>0$ there is either $\Delta=0$, $Q^2 \leq 1/(4\Lambda)$ or $\Delta>0$, $Q^2 < 1/(4\Lambda)$, and for $\Lambda<0$ there is $\Delta\leq 0$, where
\begin{equation} 
\Delta=-\frac{1}{\Lambda^5} \left(81m^4 \Lambda +Q^2 (3+4Q^2 \Lambda)-9m^2 (1+12 Q^2 \Lambda)\right).
\label{eqn:discriminant}
\end{equation}

I can get rid of the coordinate system singularities at the black hole horizons by introducing coordinates with time $t'$ defined as
\begin{equation} 
dt'=dt-\frac{\frac{2m}{r}-\frac{Q^2}{r^2}+\frac{\Lambda}{3}r^2}{1-\frac{2m}{r}+\frac{Q^2}{r^2}-\frac{\Lambda}{3}r^2}dr.
\label{eqn:tef}
\end{equation}
This yields
\begin{eqnarray}
\fl ds^2=-\left(1-\frac{2m}{r}+\frac{Q^2}{r^2}-\frac{\Lambda}{3}r^2\right) dt'^2+2\left(\frac{2m}{r}-\frac{Q^2}{r^2}+\frac{\Lambda}{3}r^2\right) dt' dr+\nonumber\\
+\left(1+\frac{2m}{r}-\frac{Q^2}{r^2}+\frac{\Lambda}{3}r^2\right) dr^2+r^2 (d\theta^2+\sin^2 (\theta) d\phi^2).
\label{eqn:metricef}
\end{eqnarray}
The determinant of the metric (\ref{eqn:metricef}) is $g=-r^4 \sin^2\theta$.

\section{Flow in Reissner--Nordstr\"{o}m--(anti--)de Sitter spacetime}
In this article I consider a perfect fluid characterized by the energy-stress tensor
\begin{equation} 
T^{\mu \nu}=(e+p)u^{\mu} u^{\nu}+p g^{\mu \nu},
\label{eqn:tensor}
\end{equation}
where $e$ denotes the energy density, $p$ is the pressure and $u^{\mu}$ is the four-velocity of the fluid. The Bondi-type (or Michel-type \cite{Bondi, Michel}) accretion is a steady, spherically symmetrical flow. In both coordinate systems, all quantities should be functions of radius only, and $u^{\theta}=u^{\phi}=0$. From the normalization of the four-velocity vector I get
\begin{eqnarray}
		u^{t'}=\frac{\left(\frac{2m}{r}-\frac{Q^2}{r^2}+\frac{\Lambda}{3}r^2\right) u^{r}+\sqrt{1-\frac{2m}{r}+\frac{Q^2}{r^2}-\frac{\Lambda}{3}r^2+\left(u^{r}\right)^{2}}}{1-\frac{2m}{r}+\frac{Q^2}{r^2}-\frac{\Lambda}{3}r^2},\label{eqn:utup}\\
		u_{t'}=-\sqrt{1-\frac{2m}{r}+\frac{Q^2}{r^2}-\frac{\Lambda}{3}r^2+\left(u^{r}\right)^{2}}.\label{eqn:utdown}
\end{eqnarray}
One can check, that in both coordinate sytems $u_{t}$ and $u_{t'}$ has exactly the same form. I will describe the motion of fluid using two conservation equations:
\begin{eqnarray}
		\nabla_{\mu}(\rho u^{\mu})=0,\label{eqn:consflow}\\
        \nabla_{\mu}((e+p)u^{\mu}u^{\nu}+pg^{\mu\nu})=0.\label{eqn:consenergy}
\end{eqnarray}
where $\rho$ denotes the baryonic (rest-mass) density. Equation (\ref{eqn:consflow}) can be rewritten as 
\begin{equation} 
\nabla_{\mu}(\rho u^{\mu})=\frac{1}{\sqrt{-g}} \partial_{\mu} (\sqrt{-g} \rho u^{\mu})=\frac{1}{r^2} \partial_{r} (r^2 \rho u^{r})=0.
\label{eqn:consflow1}
\end{equation}

Let me introduce a specific enthalpy $h=(e+p)/\rho$. Assuming that the flow is smooth, equation (\ref{eqn:consenergy}) can be expressed as
\begin{eqnarray}
\fl \nabla_{\mu}T^{\mu\nu}=\nabla_{\mu}(h \rho u^{\mu}u^{\nu})+g^{\mu\nu} \partial_{\mu}p=\rho u^{\mu}\nabla_{\mu}(h u^{\nu})+h u^{\nu} \nabla_{\mu}(\rho u^{\mu})+g^{\mu\nu} \partial_{\mu}p= \nonumber\\
=\rho u^{\mu}\nabla_{\mu}(h u^{\nu})+g^{\mu\nu} \partial_{\mu}p=0,
\label{eqn:consenergy1}
\end{eqnarray}
where I used equation (\ref{eqn:consflow}). In the following I will consider isentropic flows, for which $dh=dp/\rho$, and in consequence
\begin{equation} 
u^{\mu}\nabla_{\mu}(h u_{\nu})+\partial_{\nu}h=u^{\mu}\partial_{\mu}(h u_{\nu})-\Gamma^{\lambda}_{\nu \mu} h u_{\lambda} u^{\mu}+\partial_{\nu}h=0.
\label{eqn:consenergy2}
\end{equation}
Taking $\nu=t'$ one may find out, that the terms containing Christoffel symbols cancel. This yields
\begin{equation} 
\partial_{r}(h u_{t'})=0.
\label{eqn:consenergy3}
\end{equation}
Note that equations (\ref{eqn:consflow}), (\ref{eqn:consenergy}) may be rewritten as equations (\ref{eqn:consflow1}), (\ref{eqn:consenergy3}). Let me point out, that in both of the considered coordinate systems these equations have the same form, and consequently the obtained functions $u_t(r)$ ($u_{t'}(r)$) and $u^r(r)$ are the same in both systems. For this reason I will not differentiate between $u_t$ and $u_{t'}$ in what follows.

By integrating equations (\ref{eqn:consflow1}), (\ref{eqn:consenergy3}), and substituting $u_{t'}$ with equation (\ref{eqn:utdown}) one gets
\begin{eqnarray}
		r^2 \rho u^r=\mbox{const},\label{eqn:consflow2}\\
        h\sqrt{1-\frac{2m}{r}+\frac{Q^2}{r^2}-\frac{\Lambda}{3}r^2+\left(u^{r}\right)^{2}}=\mbox{const}.\label{eqn:consenergy4}
\end{eqnarray}
These equations constitute the starting point of the remaining part of this work.

\subsection{Sonic points}
Let $a$ denote the local speed of sound. A location in which the four-velocity of the fluid satisfies $a^2=(u^r/u_t)^2$ will be called a sonic point (because of the spherical symmetry of the flow, it is actually a sphere). I will consider barotropic equations of state ($h=h(\rho)$), so
\begin{equation} 
\frac{dh}{h}=a^2 \frac{d\rho}{\rho}.
\label{eqn:sonic1}
\end{equation}
Taking into account equation (\ref{eqn:sonic1}), differentiating equations (\ref{eqn:consflow2}) and (\ref{eqn:consenergy4}) yields
\begin{equation} 
\left[ \left(\frac{u^r}{u_t} \right)^2 -a^2 \right] \partial_r \ln u^r=\frac{1}{r (u_t)^2} \left(2a^2 (u_t)^2-\frac{m}{r}+\frac{Q^2}{r^2}+\frac{\Lambda}{3} r^2 \right).
\label{eqn:sonic2}
\end{equation}
For a sonic point (if only $\partial_r \ln u^r < \infty$) one has
\begin{equation} 
2a^2_\ast (u_{t\ast})^2-\frac{m}{r_\ast}+\frac{Q^2}{r^2_\ast}+\frac{\Lambda}{3} r^2_\ast=0.
\label{eqn:sonic3}
\end{equation}
From now on quantities with an asterisk will refer to the sonic point. Equation (\ref{eqn:sonic3}) can be rewritten as
\begin{equation}
(u^{r}_{\ast})^2=\frac{m}{2r_{\ast}}-\frac{Q^2}{2r^{2}_{\ast}}-\frac{\Lambda}{6}r^{2}_{\ast}=a^2_\ast \left(1-\frac{3m}{2r_{\ast}}+\frac{Q^2}{2r^{2}_{\ast}}-\frac{\Lambda}{2}r^{2}_{\ast}\right).
\label{eqn:sonic4}
\end{equation}

Equations (\ref{eqn:consflow2}), (\ref{eqn:consenergy4}) and (\ref{eqn:sonic1}) have to be supplemented with suitable boundary conditions. In the remaining part of the paper the values of the density and the speed of sound at the boundary of the cloud are denoted by $\rho_\infty$ and $a_\infty$. The radius of the cloud is denoted by $r_\infty$. It can be either finite or infinite depending on the context.

\subsection{Hamiltonian dynamical system}
Differentiating equations (\ref{eqn:consflow2}) and (\ref{eqn:consenergy4}) with respect to $r$, and using relation equation (\ref{eqn:sonic1}) leads to
\begin{equation}
\frac{d u^r}{dr}=\frac{2u^r}{r}\frac{a^2\left(1-\frac{2m}{r}+\frac{Q^2}{r^2}-\frac{\Lambda}{3}r^2 + \left( u^r \right)^2 \right)-\left(\frac{m}{2r}-\frac{Q^2}{2r^2}-\frac{\Lambda}{6}r^2\right)}{\left( u^r \right)^2-a^2 \left(1-\frac{2m}{r}+\frac{Q^2}{r^2}-\frac{\Lambda}{3}r^2+ \left( u^r \right)^2 \right)}.
\label{eqn:dynamical}
\end{equation}
One may introduce a parameter $l$, such that
\begin{eqnarray}
\fl
		\frac{dr}{dl}=r\left[\left(u^r\right)^2 -a^2 \left(1-\frac{2m}{r}+\frac{Q^2}{r^2}-\frac{\Lambda}{3}r^2+ \left( u^r \right)^2 \right)\right],\label{eqn:dynr} \\
\fl \frac{d u^r}{dl}=2u^r \left[a^2\left(1-\frac{2m}{r}+\frac{Q^2}{r^2}-\frac{\Lambda}{3}r^2 + \left( u^r \right)^2 \right)-\left(\frac{m}{2r}-\frac{Q^2}{2r^2}-\frac{\Lambda}{6}r^2\right)\right]. \label{eqn:dynur}
\end{eqnarray}
These equations constitute an autonomous, hamiltonian two-dimensional dynamical system. Its orbits are composed of the solutions of equations (\ref{eqn:consflow2}) and (\ref{eqn:consenergy4}). The parameter $l$ is arbitrary, so the reparametrization of this system shall not change its orbits. From equation (\ref{eqn:sonic4}) it is clear, that sonic points are the critical points of this dynamical system.

One may treat left-hand side of the equation (\ref{eqn:consenergy4}) as a hamiltonian $H$ of this system. Equation (\ref{eqn:dynamical}) may be recreated from this hamiltonian as
\begin{eqnarray}
		\frac{dr}{d\widetilde{l}}=\frac{\partial H}{\partial u^r},\label{eqn:hamr}  \\
		\frac{d u^r}{d\widetilde{l}}=-\frac{\partial H}{\partial r},\label{eqn:hamur}
\end{eqnarray}
although parameters $\widetilde{l}$ and $l$ differ.

Plots in this paper will be given in $(u^r/u_t)^2$ vs. $r$ variables, as they present information in more readable way. One may consider the above dynamical system in $u^r/u_t$ and $r$ variables. Then it will be described by the hamiltonian
\begin{equation} 
H=h\sqrt{\frac{1-\frac{2m}{r}+\frac{Q^2}{r^2}-\frac{\Lambda}{3}r^2}{1-\left(\frac{u^r}{u_t}\right)^2}}.
\label{eqn:hamiltonian}
\end{equation}
Then equations analogous to equations (\ref{eqn:hamr}) and (\ref{eqn:hamur}) yield
\begin{equation}
\frac{d\left(\frac{u^r}{u_t}\right)}{dr}=\frac{\left(\frac{u^r}{u_t}\right) \left[1-\left(\frac{u^r}{u_t}\right)^2\right]}{r\left[\left(\frac{u^r}{u_t}\right)^2-a^2\right]}\frac{2a^2\left(1-\frac{3m}{2r}+\frac{Q^2}{2r^2}-\frac{\Lambda}{2}r^2\right)-\frac{m}{r}+\frac{Q^2}{r^2}+\frac{\Lambda}{3}r^2}{1-\frac{2m}{r}+\frac{Q^2}{r^2}-\frac{\Lambda}{3}r^2}.
\label{eqn:dynamicalurut}
\end{equation}
This dynamical system has an additional critical point, which is located at the horizon.

\section{Accretion of isothermal test fluids}
In this chapter I consider the equation of state in the form $p=ke$ where $k$ is constant. In this case $a^2=dp/de$, so
$a^2=k$. Moreover, equation (\ref{eqn:sonic1}) yields
\begin{equation} 
\frac{de}{d\rho}=h=\frac{e+p(e)}{\rho}.
\label{eqn:iso1}
\end{equation}
By integrating this term from $r_\infty$ to any point inside ball of fluid I get
\begin{equation} 
\rho=\rho_\infty \exp\left(\int_{e_\infty}^{e} \frac{de'}{e'+p(e')} \right)=\rho_\infty \left(\frac{e}{e_\infty}\right)^{1/(k+1)},
\label{eqn:iso2}
\end{equation}
where the last equality assumes the equation of state $p=ke$. Combining equations (\ref{eqn:iso1}) and (\ref{eqn:iso2}) yields the following expression for the enthalpy
\begin{equation} 
h=\frac{(k+1)e_\infty}{\rho^{k+1}_{\infty}}\rho^{k}.
\label{eqn:iso3}
\end{equation}
Equation (\ref{eqn:consenergy4}) can be now written as
\begin{equation} 
\rho^k \sqrt{1-\frac{2m}{r}+\frac{Q^2}{r^2}-\frac{\Lambda}{3}r^2+\left(u^{r}\right)^{2}}=\mbox{const}.
\label{eqn:iso4}
\end{equation}
Together with equation (\ref{eqn:consflow2}) this leads to
\begin{equation} 
\sqrt{1-\frac{2m}{r}+\frac{Q^2}{r^2}-\frac{\Lambda}{3}r^2+\left(u^{r}\right)^{2}}=C r^{2k} (u^r)^k.
\label{eqn:iso5}
\end{equation}

In the following I will be especially interested in the flows that pass through the sonic point, although I also describe non-transonic solutions. Transonic flows are interesting for many reasons. They are often considered in a context of spherical accretion \cite{karkowski_malec, Mach, Bondi, Michel}. Their physical meaning comes from the fact that they maximize accretion rate \cite{Bondi, Malec} and are global in some cases \cite{Bondi}. This second property also holds for some types of spacetimes and matter considered in this paper.

For $a^2=k$ equation (\ref{eqn:sonic4}) can be written as
\begin{equation} 
\frac{m}{2r_{\ast}}-\frac{Q^2}{2r^{2}_{\ast}}-\frac{\Lambda}{6}r^{2}_{\ast}=k \left(1-\frac{3m}{2r_{\ast}}+\frac{Q^2}{2r^{2}_{\ast}}-\frac{\Lambda}{2}r^{2}_{\ast}\right).
\label{eqn:iso6}
\end{equation}
Equation (\ref{eqn:iso5}) can be solved numerically for any $k$. On the other hand for $k=1, \frac{3}{4}, \frac{2}{3}, \frac{1}{2}, \frac{1}{3}, \frac{1}{4}$ it reduces to at most four degree polynomial equations of $u^r(r)$ and can be solved analitically. In this paper I will consider four cases: $k=1$ (ultra-stiff fluid), $k=\frac{1}{2}$ (ultrarelativistic fluid), $k=\frac{1}{3}$ (radiation fluid), and $k=\frac{1}{4}$ (subrelativistic fluid).

\subsection{Solution for $k=1$}
A fluid obeying the equation of state $p=ke$ for $k=1$ is called the ultra-stiff fluid. In this case equation (\ref{eqn:iso5}) implies that 
\begin{equation} 
\left(u^r\right)^2=\frac{1-\frac{2m}{r}+\frac{Q^2}{r^2}-\frac{\Lambda}{3} r^2}{C r^4-1}.
\label{eqn:k111}
\end{equation}
There exist two solutions $u^r(r)$ differing only in sign. The expression for $r_\ast$ is much more complicated. Equation (\ref{eqn:iso6}) is identical to the expression for the locations of the horizons (\ref{eqn:singular}) so the sonic point has to be located on the horizon. The solution can be obtained analytically, though the resulting expressions are lengthy. Because of that, I will not provide explicit solutions.

Knowing $r_\ast$ allows one to compute $u^r_\ast$ from equation (\ref{eqn:sonic4}). These two numbers allow one to obtain the constant $C$ in equation (\ref{eqn:k111}); that results in the explicit form of the function $u^r(r)$. Finally it is possible to get $u_t(r)$ from equation (\ref{eqn:utdown}). Sample plots of $(u^r / u_t)^2(r)$ for this and other cases are shown in figures \ref{fig:izoplus}, \ref{fig:izominus}.

\subsection{Solution for $k=\frac{1}{2}$}
In this case, describing ultrarelativistic fluid, equation (\ref{eqn:iso6}) results in
\begin{equation} 
\frac{\Lambda}{6}r^{4}_{\ast}-r^2_\ast+\frac{5m}{2}r_{\ast}-\frac{3Q^2}{2}=0.
\label{eqn:k121}
\end{equation}
It is also quartic, similarly to the $k=1$ case, therefore I will not provide the explicit solution. On the other hand equation (\ref{eqn:iso5}) becomes a quadratic equation with the solution of the form
\begin{equation}
u^r=-\frac{1}{2}C r^2 \pm \frac{1}{2} \sqrt{C^2 r^4 -4 \left(1-\frac{2m}{r}+\frac{Q^2}{r^2}-\frac{\Lambda}{3} r^2 \right)}.
\label{eqn:k122}
\end{equation}
Again one can obtain $r_\ast$ and $u^r_\ast$ from equations (\ref{eqn:k121}) and (\ref{eqn:sonic4}), and then use them to calculate $C$ from equation (\ref{eqn:k122}), repeating procedure from the previous paragraph.

With two solutions (\ref{eqn:k122}), for a generic set of parameters I get two possible functions $(u^r / u_t)^2$ describing fluids passing through the sonic point. This corresponds to standard nonrelativistic accretion considered by Bondi in \cite{Bondi}. The standard interpretation is that one of the mentioned functions describes gas accreting onto black hole, while the second one is connected with so-called stellar wind. Both solutions are shown in figures \ref{fig:izoplus}, \ref{fig:izominus}. 

\begin{figure}[h]
\includegraphics[width=0.7\textwidth]{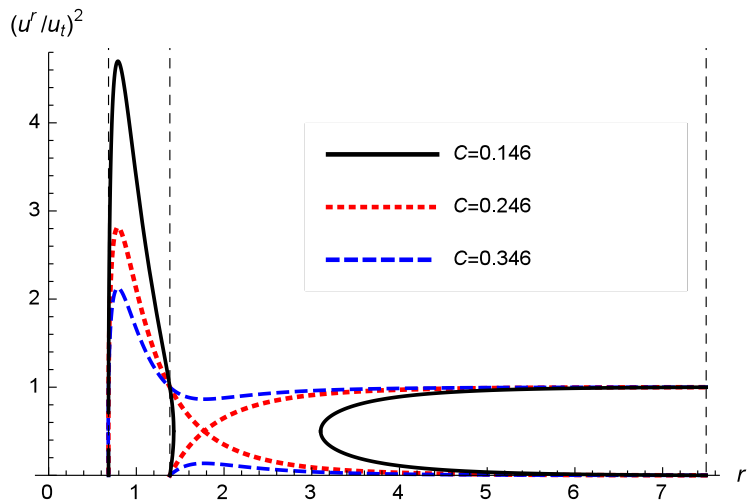}\\
\caption{Solutions obtained for isothermal equations of state $p=e/2$ with $m=1$, $Q=95/100$ and $\Lambda=1/25$ for $C$ constant (from equation (\ref{eqn:k122})) equal to 0.146, 0.246 (transonic solution), and 0.346. Dashed vertical lines denote locations of the horizons.}
\label{fig:12entropia}
\end{figure}

Small changes of the constant $C$ lead to solutions characterized by different values of entropy. An example of solutions generated for other values of $C$, and therefore not passing through the sonic point, is presented in figure \ref{fig:12entropia}.

\subsection{Solution for $k=\frac{1}{3}$}\label{sec:k13}
The case $k=\frac{1}{3}$ is especially interesting, as it describes radiation fluid. Moreover, for $k=\frac{1}{3}$ equation (\ref{eqn:iso6}) can be reduced to the quadratic equation
\begin{equation} 
r^{2}_{\ast}-3mr_{\ast}+2Q^2=0.
\label{eqn:k131}
\end{equation}
The critical points are located at 
\begin{equation} 
r_{\ast}=\frac{3m \pm \sqrt{9m^2-8Q^2}}{2}.
\label{eqn:k132}
\end{equation}
Please note, that the location of the critical point is independent of the value of the cosmological constant. The same effect can be observed for the Schwarzschild-(anti--)de Sitter spacetime \cite{Mach}. As it can be seen in figures \ref{fig:izoplus}, \ref{fig:izominus} one of these two points is a saddle point, when the second one seems to be a center point.

The relation between radial component of the four-velocity $u^r$ and $r$ can be described as
\begin{equation} 
\left(1-\frac{2m}{r}+\frac{Q^2}{r^2}-\frac{\Lambda}{3}r^2+\left(u^{r}\right)^{2}\right)^3=C r^{4} (u^r)^2.
\label{eqn:k133}
\end{equation}
As before one can calculate $(u^r_\ast)^2$,
\begin{eqnarray} 
\fl (u^r_\ast)^2=\frac{\pm 3 m \sqrt{9 m^2-8 Q^2} \left(1-4 \Lambda  Q^2\right)-9 m^2 \left(4 \Lambda  Q^2+1\right)+4 Q^2 \left(4 \Lambda  Q^2+3\right)}{48 Q^2}.
\label{eqn:k134}
\end{eqnarray}
Inserting equations (\ref{eqn:k132}) and (\ref{eqn:k134}) into equation (\ref{eqn:k133}) allows one to find the value of the constant $C$ depending on $m$, $\Lambda$, and $Q$. Varying the value of $C$ leads to results similar to the ones for the $k=\frac{1}{2}$ case.

\begin{figure}[h]
\includegraphics[width=\textwidth]{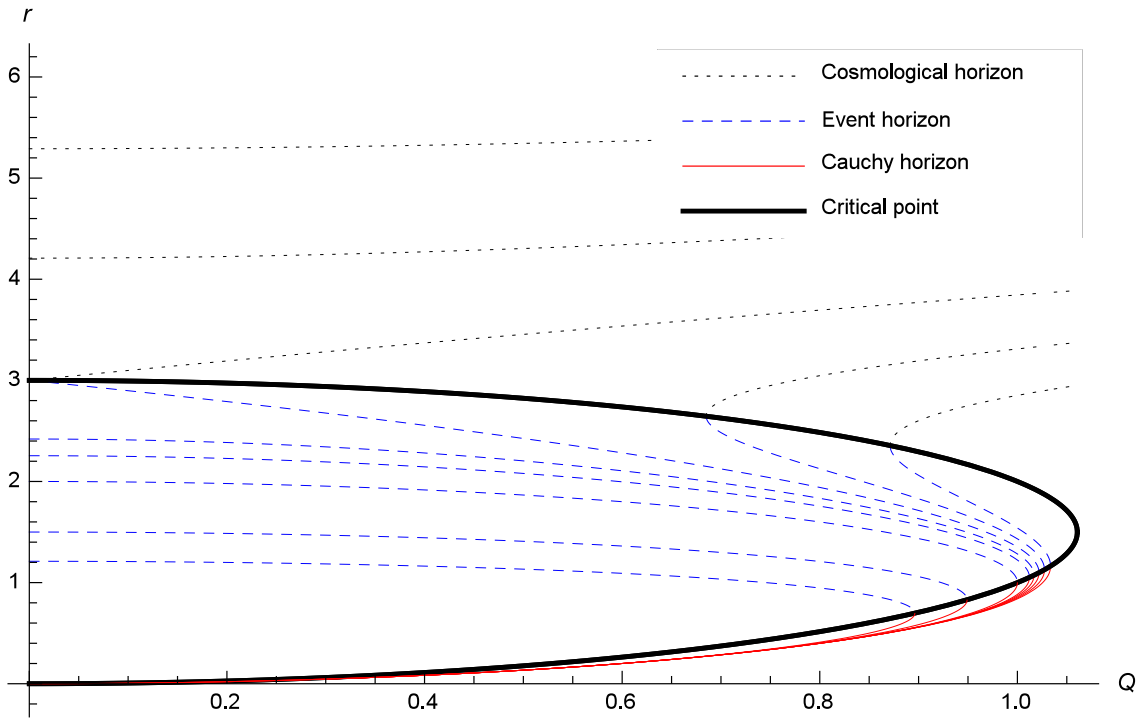}\\
\caption{Locations of the horizons and the critical points depending on the cosmological constant $\Lambda$ and the black hole's charge $Q$ for the equation of state in a form $p=e/3$. Solid, red lines denote Cauchy horizons, dashed, blue lines denote event horizons, dotted, black lines denote cosmological horizons and a thick, black line denotes critical points. Different lines correspond to cases with different values of the cosmological constant $\Lambda$ (changing from -16/90 to 1/5). Location of the critical points does not depend on the cosmological constant.}
\label{fig:horizons}
\end{figure}

There exists a correlation between the locations of the critical points for $k=\frac{1}{3}$ and black hole's horizons, which is shown in figure \ref{fig:horizons}. It can be proven, that if there exist three horizons, one of them must be above both critical points, one of them must be between the critical points, and one must be below both critical points. For $\Lambda<0$, when there exist two horizons, one is below both critical points and one is located between them. The proof can be found in the Appendix. This property implies, that the horizons of two types can overlap each other only in the critical points.

\subsection{Solution for $k=\frac{1}{4}$}
When $k=\frac{1}{4}$, equations (\ref{eqn:iso6}) and (\ref{eqn:iso5}) yield
\begin{equation} 
\frac{\Lambda}{6}r^{4}_{\ast}+r^2_\ast-\frac{7m}{2}r_{\ast}+\frac{5Q^2}{2}=0,
\label{eqn:k141}
\end{equation}
\begin{equation} 
\left(1-\frac{2m}{r}+\frac{Q^2}{r^2}-\frac{\Lambda}{3}r^2+\left(u^{r}\right)^{2}\right)^2=C r^{2} u^r.
\label{eqn:k142}
\end{equation}
By repeating the procedure from previous paragraphs one can obtain the function $(u^r / u_t)^2(r)$. Again varying the constant $C$ in  equation (\ref{eqn:k142}) results with solutions that are similar to those obtained in the previous case.

\begin{figure}[h]
\includegraphics[scale=1]{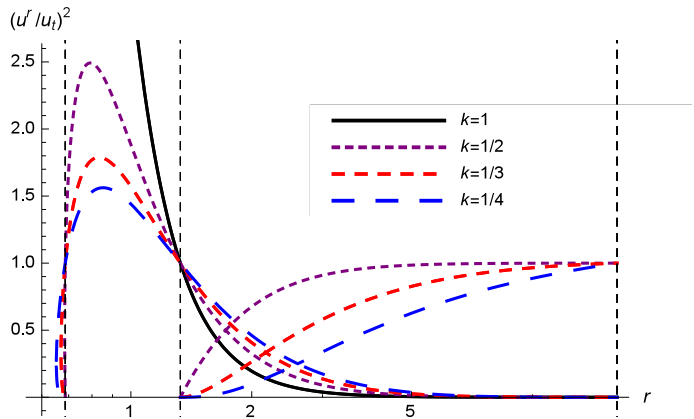}\\
\caption{Transonic solutions obtained for the isothermal equation of state $p=ke$ with $m=1$, $Q=95/100$ and $\Lambda=1/100$. Dashed vertical lines denote locations of the horizons.}
\label{fig:izoplus}
\end{figure}

\begin{figure}[h]
\includegraphics[scale=1]{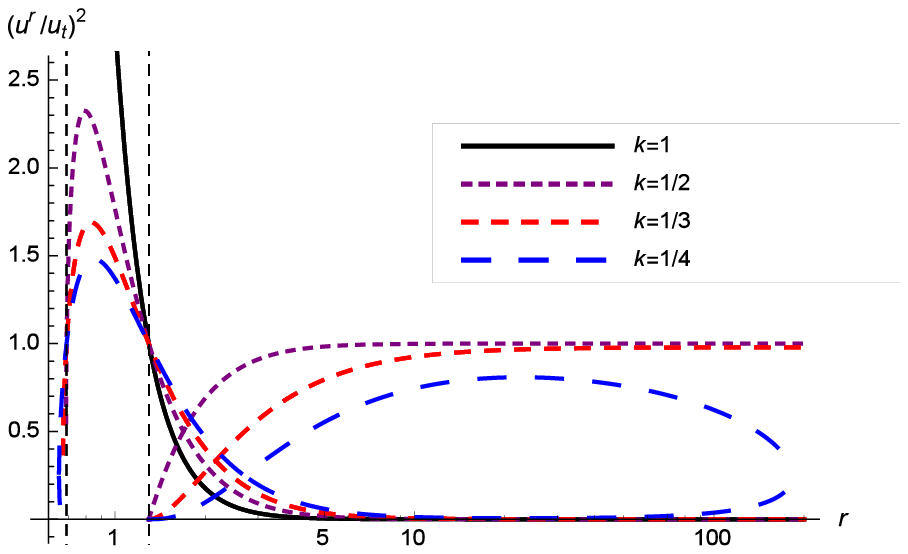}\\
\caption{Transonic solutions obtained for the isothermal equation of state $p=ke$ with $m=1$, $Q=95/100$ and $\Lambda=-1/100$. Dashed vertical lines denote locations of the horizons.}
\label{fig:izominus}
\end{figure}

The solutions are showed in figures \ref{fig:izoplus}, \ref{fig:izominus}. One can observe, that in this case it is possible to get a closed trajectory on a plot $(u^r/u_t)^2$ vs. $r$, as it is depicted in figure \ref{fig:izominus}. Further discussion can be found at the end of the next section.

\section{Accretion of polytropic test fluids}
\begin{figure}[h]
\includegraphics[width=\textwidth]{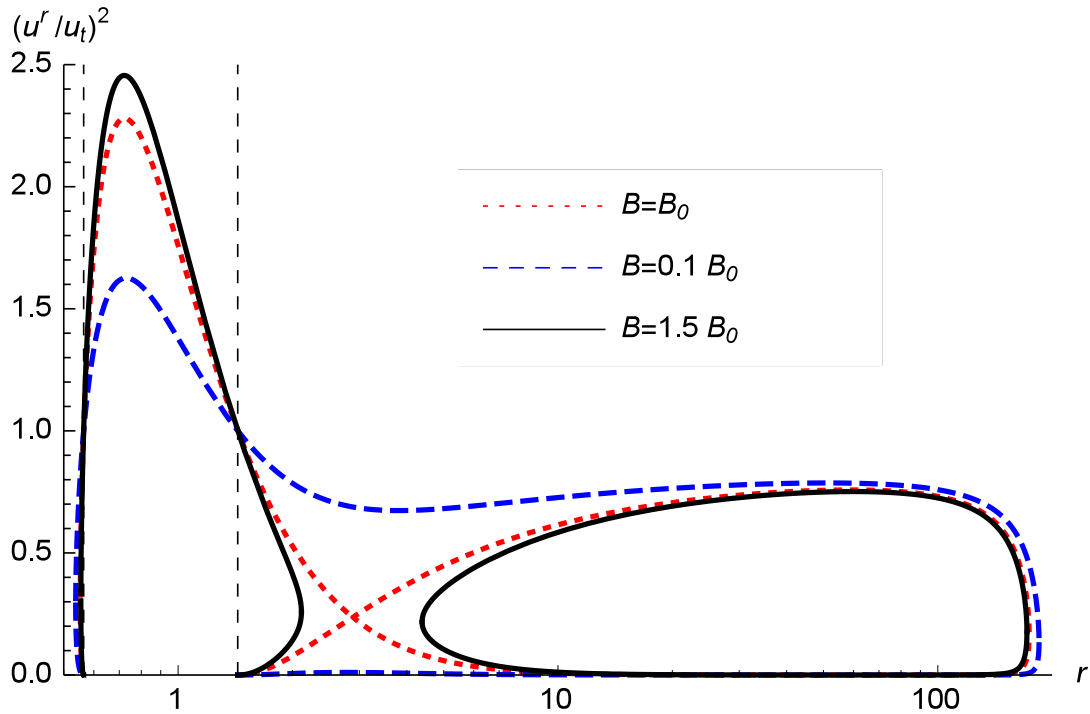}\\
\caption{Solutions for polytropic flow characterized by $m=1$, $Q=9/10$, $\Lambda=-3.535\times10^{-4}$, $\Gamma=4/3$, $r_\infty=10^6$, and $a^2_\infty=0.2$. The plot presents transonic solution (obtained for $B=B_0$) and two non-transonic.}
\label{fig:poli}
\end{figure}

In this section I assume a polytropic equation of state of the form $p=K \rho^\Gamma$, where $K$ and $\Gamma$ are constants. For this equation of state
\begin{equation} 
h=\frac{\Gamma-1}{\Gamma-1-a^2}.
\label{eqn:poli1}
\end{equation}
Together with equation (\ref{eqn:consenergy4}) it gives
\begin{equation} 
\frac{\sqrt{1-\frac{2m}{r}+\frac{Q^2}{r^2}-\frac{\Lambda}{3}r^2+\left(u^{r}\right)^{2}}}{\Gamma-1-a^2}=\mbox{const},
\label{eqn:poli2a}
\end{equation}
and in particular
\begin{eqnarray}
(\Gamma-1-a^2)\sqrt{1-\frac{2m}{r_\infty}+\frac{Q^2}{r^2_\infty}-\frac{\Lambda}{3}r^2_\infty+\left(u^{r}_\infty\right)^{2}}=\nonumber\\
=(\Gamma-1-a^2_\infty)\sqrt{1-\frac{2m}{r}+\frac{Q^2}{r^2}-\frac{\Lambda}{3}r^2+\left(u^{r}\right)^{2}}.
\label{eqn:poli2}
\end{eqnarray}
Using equation (\ref{eqn:sonic1}) one can find, that for the polytropic equation of state
\begin{equation} 
\rho=\rho_\infty \left(\frac{a^2}{a^2_\infty} \frac{\Gamma-1-a^2_\infty}{\Gamma-1-a^2}\right)^\frac{1}{\Gamma-1}.
\label{eqn:poli3}
\end{equation}
This, in connection with equation (\ref{eqn:consflow2}), yields
\begin{equation} 
u^r=u^r_\infty \frac{r^2_\infty}{r^2} \left(\frac{a^2_\infty}{a^2} \frac{\Gamma-1-a^2}{\Gamma-1-a^2_\infty}\right)^\frac{1}{\Gamma-1}.
\label{eqn:poli4}
\end{equation}
Combining equations (\ref{eqn:poli2}) and (\ref{eqn:poli4}) one gets
\begin{eqnarray}
\fl (\Gamma-1-a^2_\ast)\sqrt{1-\frac{2m}{r_\infty}+\frac{Q^2}{r^2_\infty}-\frac{\Lambda}{3}r^2_\infty+B}=\nonumber\\
=(\Gamma-1-a^2_\infty)\sqrt{1-\frac{3m}{2r_\ast}+\frac{Q^2}{2r_\ast^2}-\frac{\Lambda}{2}r^2_\ast+\left(u^r_\ast \right)^2},
\label{eqn:poli5}
\end{eqnarray}
where
\begin{equation} 
B=\left(u^{r}_\ast\right)^{2} \frac{r^4_\ast}{r^4_\infty} \left(\frac{a^2_\ast}{a^2_\infty} \frac{\Gamma-1-a^2_\infty}{\Gamma-1-a^2_\ast}\right)^\frac{2}{\Gamma-1}.
\label{eqn:poli5}
\end{equation}
Substituting $a_\ast^2$ and $u^{r}_\ast$ with formulas taken from equation (\ref{eqn:sonic4}), one gets equation (\ref{eqn:poli5}) with $r_\ast$ as the only one unknown. Solving this equation with boundary values of $r_\infty$ and $a_\infty$ gives positions of the critical points $r_\ast$ and, as a result, values of $u_\ast$, $a^2_\ast$. The knowledge of $r_\infty$, $a_\infty$, $r_\ast$, $u_\ast$, and $a^2_\ast$ allows one to determine $u^r_\infty$ from equation (\ref{eqn:poli4}). Finally one can calculate the constant in equation (\ref{eqn:poli2a}), at $r_\infty$. Combining this formula with equation (\ref{eqn:poli4}) provides an implicit formula for $u^r(r)$. Solving it numerically and using equation (\ref{eqn:utdown}) one can get the function $(u^r/u_t)^2 (r)$. Similiary to the case of the isothermal fluid one may variate the constant $B$ from equation (\ref{eqn:poli5}) obtaining solutions that are not transonic. Figure \ref{fig:poli} shows a sample plot of the function $(u^r/u_t)^2 (r)$. 

Figure \ref{fig:poli} demonstrates, that it is possible to get a closed, subsonic solution. One can obtain a solution in a binocular shape consisting of two closed curves --- one above the sonic point and one below it, laying partially under the event horizon. The transonic solution is also closed and similar to one obtained for isothermal fluid with $k=\frac{1}{4}$. It has been known, that in the case of the Reissner-Nordstr\"{o}m spacetime one can obtain flow trajectories closed near the Cauchy and event horizons \cite{Babichev}. On the other hand, the possibility of getting trajectories closing for radii far from the event horizon in the Schwarzshild-anti-de Sitter spacetime has been revealed in \cite{Mach}. It suggests, that in the presence of the both cosmological constant and charge it should be possible to get trajectories closed from both sides. The results described here confirm that in special cases of the subrelativistic isothermal flow or polytropic flow. This is especially interesting in the context of rotating black holes. As it was mentioned in the Introduction, there exist qualitative similarities between Reissner-Nordstr\"{o}m spacetime and the Kerr spacetime. For example one may apply swiss cheese model to a charged black hole in order to approximate real astrophysical situation \cite{Einstein}. The obtained results reveal interesting properties of the flows in such a situation.

\section{Summary}
In this paper I considered isothermal and polytropic flows in the Reissner--Nordstr\"{o}m--(anti--)de Sitter spacetime. I described the way to find sonic points and to obtain the solutions $(u^r/u_t)^2 (r)$. One may think about these sonic points as critical points of a hamiltonian dynamical system connected with this flow. I investigated specific cases of isothermal test fluids: so-called ultra-stiff fluid, ultrarelativistic fluid, radiation fluid, and subrelativistic fluid. In these cases the problem of solving the equations of motion of the fluid can be reduced to solving a polynomial of at most four degree. Exemplary plots of the flows were showed. I proved the correlation between positions of the horizons and the sonic points for the radiation fluid. Polytropic flows were also studied. It occured that in case of the subrelativistic isothermal flow or polytropic flow it is possible to obtain closed, binocular-like trajectories in $(u^r/u_t)^2 (r)$ phase space. This property is consistent with previous results considering spacetimes with either non-zero cosmological constant \cite{Mach} or charge \cite{Babichev}, where trajectories were closing respectively in the neighbourhood of the cosmological horizon and the Cauchy horizon.

This is especially interesting in the context of rotating black holes. As it was mentioned in the Introduction, there exist qualitative similiarities between Reissner--Nordstr\"{o}m spacetime and the Kerr spacetime. The obtained results reveal interesting properties of the accretion in the Reissner--Nordstr\"{o}m case. I believe that at least some of them can be present in Bondi-type accretion on rotating black holes.

This paper fits into current research on closed flow trajectories (cf. \cite{Mach, mach_sam, Babichev, Chaverra}). It appears, that whether solutions of this type exist, depends on both the spacetime and the equation of state of the fluid. In the case of the Reissner--Nordstr\"{o}m--anti--de Sitter spacetime it is relatively easy to obtain such solutions --- they appear to exist for matter models that can be interpreted as a gas of massive particles \cite{mach_sam}. In other types of spacetimes it may be not so common, for example for a pure Schwarzschild solution one may have closed trajectories in more exotic case of polytropes with a high polytropic index \cite{Chaverra}. The dependence of the existence of the closed solutions of considered spacetime and fluid types is a problem that needs further investigation. Of course it would be especially enlightening to compare results obtained here to corresponding results for Kerr spacetimes, which unfortunately do not exist so far.

\ack
Primarily I thank Dr. Patryk Mach for the assistance. Without his endless aid, long dicussions, and explanations, this paper would never come into existence. Also some numerical work was based on his code. I thank Jerzy Knopik for productive talks, which sometimes were crucial for understanding. At the end I thank Prof. Edward Malec for important remarks.

\appendix
\setcounter{section}{1} 
\section*{Appendix}
In this appendix I prove the relation between locations of the horizons and critical points that was discussed in Section \ref{sec:k13}. In the generic case of the Reissner--Nordstr\"{o}m--de Sitter spacetime, when there exist three horizons, one of them is always located above both critical points, one is between the critical points, and one is below both critical points. Similiar property also holds for the Reissner--Nordstr\"{o}m--anti-de Sitter spacetime: when there exist two horizons, one of them is between the critical points, and the second one is below both critical points. 

\begin{figure}[h]
\includegraphics[width=\textwidth]{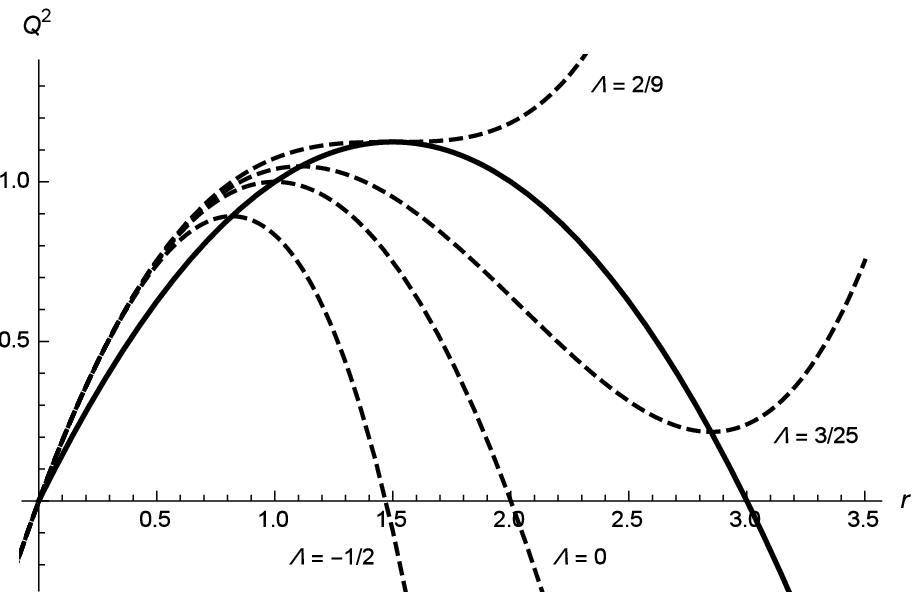}\\
\caption{Plots of $f(r)$ (dashed) and $g(r)$ (solid) for $m=1$ and sample $\Lambda$ values.}
\label{fig:proof}
\end{figure}

In order to prove it, let me consider two polynomials:
\begin{eqnarray}
		f(r)=\frac{\Lambda}{3} r^4 - r^2 + 2 m r,\label{eqn:proof1a}\\
        g(r)= -\frac{1}{2} r^{2}+\frac{3}{2}mr.\label{eqn:proof1b}
\end{eqnarray}
These functions are formulas for $Q^2$ derived from the equations (\ref{eqn:singular}) and (\ref{eqn:k131}). For fixed $m$ and $\Lambda$, the function $f(r)$ yields a value of $Q^2$ for which there is a horizon in $r$. Similiary the function $g(r)$ yields $Q^2$ for which there is a critical point in $r$. Inverses of these functions were plotted in figure \ref{fig:horizons}. I consider behaviour of these polynomials for $r>0$ and $Q^2>0$. The function $f(r)$, as a quartic function, may have at most three extrema. The function $g(r)$ is a quadratic function with a maximum at $r=3m/2$. Sample plots of functions $f(r)$ and $g(r)$ are presented in figure \ref{fig:proof}. As I demand the existence of three horizons in $\Lambda>0$ case and two horizons in $\Lambda<0$ case, I am interested in situations when for $r>0$ and a fixed value of $Q^2$ there are respectively three and two values of $r$ such that $f(r)=Q^2$.

By differentiating $f(r)$ one can obtain a condition for extrema of $f$:
\begin{equation} 
f'(r)=\frac{4\Lambda}{3} r^3-2r+2m=0.
\label{eqn:proof3}
\end{equation}
The case when an extremum is an inflection point ($f''(r)=0$) will be considered later. Substracting equations (\ref{eqn:proof1a} and \ref{eqn:proof1b}) results in
\begin{equation} 
f(r)-g(r)=\frac{r}{4}\left(\frac{4\Lambda}{3} r^3-2r+2m\right).
\label{eqn:proof4}
\end{equation}
It implies, that for $r\neq 0$ functions $f(r)$ and $g(r)$ have $f(r)=g(r)$ at extrema of $f(r)$. Both functions $f(r)$ and $g(r)$ vanish at $r=0$. Moreover one has $f'(0)=2m>3m/2=g'(0)$. Let me now consider the two cases.

At first I will assume $\Lambda>0$. I demand the existence of three horizons, so for $r>0$ the function $f(r)$ has to have two extrema (it is clear that it cannot have more extrema, as $f'(0)>0$). Beacause $f'(0)>g'(0)>0$ and $\Lambda>0$ the first of these extrema will be a maximum (at $r_1$) and the second one will be a minimum (at $r_2$). Of course $f(r_1)=g(r_1)$ and $f(r_2)=g(r_2)$. It means that the function $f(r)$ is increasing for $r<r_1$. Then it is decreasing until $r$ reaches $r_2$. After that $f(r)$ goes to infinity. Three horizons can exist only, if a fixed $Q^2$ (such that for the horizons $f(r)=Q^2$) satisfies $r_2<Q^2<r_1$. It implies, that one of the horizons is located on the left side of the parabole $g(r)$ (below the critical points), one inside it (between the critical points), and one on the right side (above the critical points).

A similiar argument works in the case of $\Lambda<0$. This time I demand the existence of two horizons in the half-plane $r>0$. Again one has $f'(0)>g'(0)>0$. Beacause $\lim_{r\to\infty} f(r) = -\infty$, the graph $f(r)$ has to cross the graph $g(r)$ in a point $r$ such that $0<r<3m/2$. Then $f(r)$ starts decreasing and cannot cross $g(r)$ again. In the opposite situation it would have a minimum and then would start increasing, while $g(r)$ would be still decreasing. The function $f(r)$ could not change the sign of derivative and one would have $\lim_{r\to\infty} f(r) = \infty$, what stays in a contradiction with the assumption that $\Lambda<0$. It means, that there exists at most two horizons --- one below both critical points and one between them.

The function $f(r)$ does not have an extremum in the point described by equation (\ref{eqn:proof3}), if additionally $f''(r)=0$ in the same point. Both of these conditions lead to $\Lambda=2/(9m^2)$ and the occurence of a inflection point at $r=3m/2$ (the maximum of the function $g(r)$). Obviously this is possible only in the Reissner--Nordstr\"{o}m--de Sitter spacetime. In this case there is only one horizon (which is located below both critical points).

\end{document}